\documentclass{article}
\usepackage{waspaa21}
\usepackage{cite}
\usepackage{amsmath,amssymb,amsfonts}
\usepackage{algorithmic}
\usepackage{graphicx}
\usepackage{textcomp}
\usepackage{xcolor}
\usepackage{balance}
\usepackage{caption}
\usepackage{subcaption}
\usepackage[version-1-compatibility]{siunitx}
\usepackage{multirow}
\usepackage{siunitx}
\def\BibTeX{{\rm B\kern-.05em{\sc i\kern-.025em b}\kern-.08em
    T\kern-.1667em\lower.7ex\hbox{E}\kern-.125emX}}

\title{A Universal Deep Room Acoustics Estimator}

\twoauthors
{$\text{Paula Sánchez López}^*$\thanks{*Work completed during an internship at Logitech.}, Paul Callens}
{LTS2, \'{E}cole Polytechnique F\'{e}d\'{e}rale de Lausanne,\\ 
Station 11, CH-1015 Lausanne,\\
Switzerland}
{Milos Cernak}
{Logitech Europe S.A., \\
CH-1015, Lausanne, \\
Switzerland}

\begin{document}

\ninept
\maketitle

\begin{sloppy}

\begin{abstract}
Speech audio quality is subject to degradation caused by an acoustic environment and isotropic ambient and point noises. The environment can lead to decreased speech intelligibility and loss of focus and attention by the listener. Basic acoustic parameters that characterize the environment well are (i) signal-to-noise ratio (SNR), (ii) speech transmission index, (iii) reverberation time, (iv) clarity, and (v) direct-to-reverberant ratio. Except for the SNR, these parameters are usually derived from the Room Impulse Response (RIR) measurements; however, such measurements are often not available. This work presents a universal room acoustic estimator design based on convolutional recurrent neural networks that estimate the acoustic environment measurement blindly and jointly. Our results indicate that the proposed system is robust to non-stationary signal variations and outperforms current state-of-the-art methods.

\end{abstract}


\section{Introduction}
Speech intelligibility is defined as the recognition rate of meaningful (dictionary) words given environmental conditions~\cite{fletcher1929speech,allen2005articulation}.  The environmental conditions include room acoustics effects (e.g., reverberation) and the presence of background noise~\cite{STIestimation}.  The speaker may be unable to recognize alterations on these conditions during communication (e.g., video call, remote presentation, voice recording), leading to reduced speech intelligibility, which is well-known to cause misunderstanding and loss of interest by the  listener~\cite{speechIntell}. Therefore, it is of great interest to develop a system that could estimate and provide instantaneous feedback on background noise and room acoustic effect parameters. This system could help the user improve their speech intelligibility during a performance, as attempted by some recent works~\cite{voiceAssist}.

Room acoustic effects on speech intelligibility can be measured using standard acoustic parameters \cite{BSENISO3382-1:20092009AcousticsSpacesb}: (i) \textit{Reverberation time (RT60)} is defined by the time it takes for the sound energy to decay after the source is switched off; (ii) \textit{Clarity (C50/C80)} is measured by calculating the ratio between the early reflections energy (up to 50/80ms) and the energy of the late response from the decay curve; (iii) \textit{Direct-to-Reverberant Ratio (DRR)}, which, similarly to C50, is a ratio of the direct sound energy over the later energy, considering that the direct sound is included in the leading 2.5  milliseconds of the RIR \cite{callens2020joint}; (iv) the \textit{Speech Transmission Index (STI)} is a metric ranging between  0  and  1  
that measures how the recording environment warps the modulations of speech at frequencies that are important to speech perception~\cite{stiNONlinear}. These parameters are traditionally derived from Room Impulse Response (RIR) measurements. Nevertheless, measuring the RIR is a laborious process that requires specific equipment~\cite{Farina2000SimultaneousTechnique}. Some attempts were conducted in the last decade to estimate acoustic parameters directly from audio signals. In 2018, Gamper and Tashev used Convolutional Neural Networks (CNNs) and Gammatone filterbanks to predict the average RT60 of a reverberant signal~\cite{gamper2018blind} and outperformed the best Acoustic Characterisation of Environments (ACE) challenge~\cite{eaton2016estimation} method. Seetharaman et al. made no assumptions about the model of the RIR and used deep CNNs to estimate the STI~\cite{STInew}. Recent work of Looney and Gaubitch~\cite{looney2020joint} showed promising results in the joint blind estimation of RT60, DRR, and Signal-to-Noise Ratio (SNR). We replicated and improved joint R60 and DRR estimation performance in our recent work as well~\cite{callens2020joint}.

The Signal-to-Noise Ratio (SNR) is the most popular metric to compare a signal's level to the background noise level. The SNR is defined as the ratio of signal power to noise power, often expressed in decibels ($\dB$). Estimation of the global SNR has been widely studied. Previous algorithms were based on identifying the noise and speech energy distributions, or signal statistics~\cite{SNRgauss, SNRwada}. More recent studies have focused on estimating the ideal binary mask (IBM) that segregates speech and noise units in a time-frequency representation of the noisy speech signal~\cite{CASAnew, CASA, SNRCASA, SNRIBM}. 

This work presents a deep convolutional recurrent neural network architecture that blindly estimates the mentioned room acoustic parameters (RT60, C50, C80, DRR, and STI) and the SNR of an input reverberant and noisy speech signal without comparing it to a reference ``clean`` recording. Our data augmentation pipeline simulates both room reverberation and different types of noise in contrast to previous studies. 
This model could be used in embedded systems to give speakers instantaneous feedback.
Users could benefit from this feedback to improve their presentations and recordings within other applications. We named the proposed system ``the universal acoustic room estimator``, a single model that estimates room acoustics' parameters.

Our central hypothesis of the proposed work is that room acoustics is well represented by the aforementioned complementary parameters, and multi-task ML training can benefit from it. Thus, the joint model should perform better than isolated predictors. Besides, most previous studies of blind acoustic environment characterization explored mostly CNNs~\cite{looney2020joint}. In this work, we try to find the decay of a signal, and it makes sense considering the sequence aspect of data, and thus using recurrent layers might be beneficial.

The paper is organized as follows: Section~\ref{sec:method} describes the methods: used dataset, data augmentation pipeline, and proposed model implemented. Section~\ref{sec:results} presents obtained results. Discussion and conclusion follow in Section~\ref{sec:conclusion}.


\section{Methods}
\label{sec:method}

Building upon our previous work~\cite{callens2020joint}, the proposed universal acoustic estimator is based on a convolutional recurrent neural network (CRNN) architecture that jointly estimates room acoustic parameters and the SNR. Figure~\ref{fig:pipeline} shows the pre-processing pipeline. The method consists of the following main steps: i) data preparation, ii) baseline model implementation, iii) proposed neural network implementation, and iv) evaluation of the models. 
\begin{figure*}[htbp]
   \centering
   \includegraphics[width=0.9\linewidth]{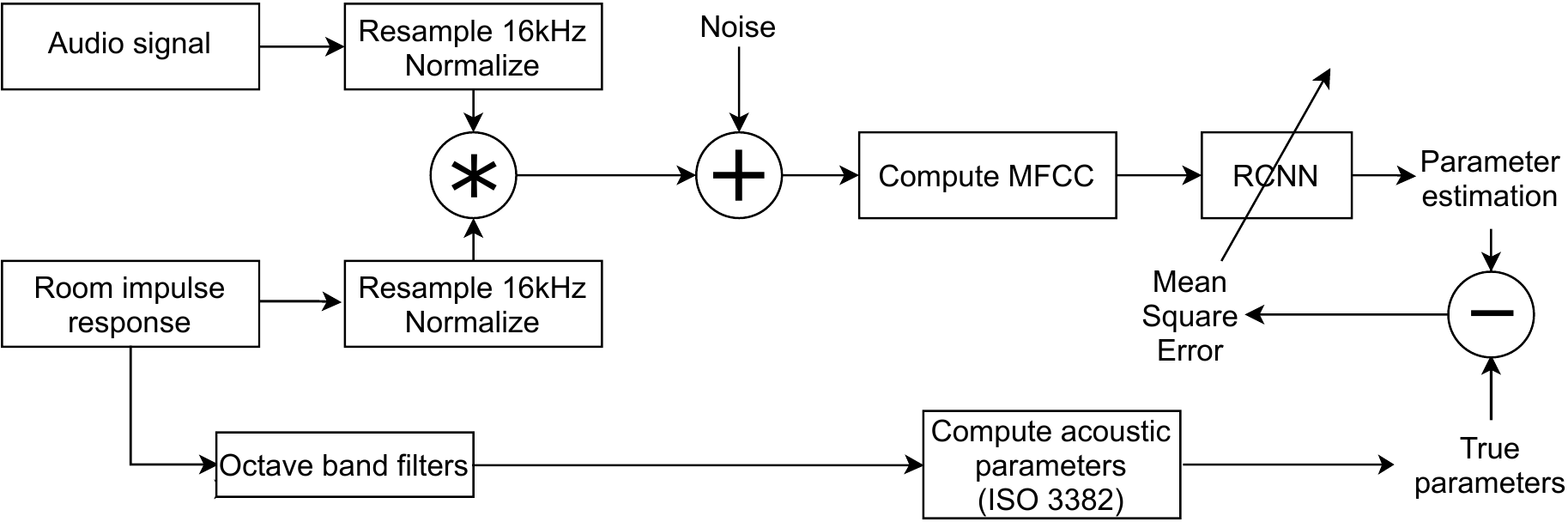}
   \caption{Overall acoustic parameter estimation pipeline. Preparation of the true parameters is described in Sec.~\ref{sec:truelabs}. The actual parameter estimation of testing audio includes only the MFCCs calculation and CRNN inference.}
   \label{fig:pipeline}\end{figure*}
   
\subsection{Data preparation}
\label{sec:truelabs}

A source sound signal $x(t)$ coming from a speaker is subject to reverberation and noise $n(t)$ when played in a room. The resulting signal can be expressed as:
\begin{equation}
    y(t) = x(t)*h(t) + n(t)
    \label{eq:1}
\end{equation}
with $h(t)$ the RIR. The training dataset and test dataset generated in this work were obtained following Equation \ref{eq:1}.  Reverberation was simulated by convolving speech with RIRs. The noise was simulated by adding real, white, or pink noise to the (non)-reverberant signal. Some speech signals did not present reverberation or background noise. 

Clean speech signals ($x(t)$) and real background noise signals ($n(t)$) were collected from the MUSAN dataset \cite{musan2015}. RIRs ($h(t)$) were gathered and sorted from online open-source libraries: OpenAirLib, echo-thief, MIT IR Survey, and RWCP. The MUSAN data was split into 80/20\% training and testing subsets, respectively. The 406 collected RIRs were split into 306 used for training and 100 for testing.

The values of true room acoustic parameters were extracted from the RIR recordings. They are computed as defined in the International Standard of Room Acoustic Measurements, ISO3382~\cite{international2009acoustics}. Figure \ref{fig:ac_distrib} shows the distribution of the measured acoustic parameters from our set of RIRs after it has been balanced. It also highlights the correlation between T60, DRR, and C50, which can be interpreted as the fact that a room with a high reverberation goes along with lower clarity and direct-to-reverberant ratio; thus, the speech will be harder to understand.

\begin{figure}[htbp]
    \centering
    \includegraphics[width=\linewidth]{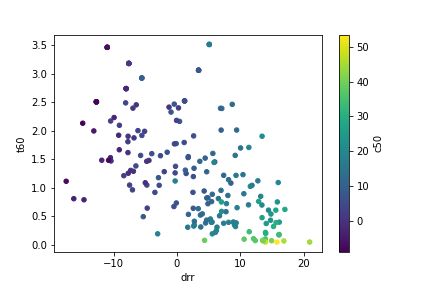}
    \caption{Distribution of the true (measured) acoustic parameters.}
    \label{fig:ac_distrib}
\end{figure}

To compute RT60, the decay curve is extracted from the RIR using Schroeder integration\cite{Schroeder1979Integrated-impulseImpulses} given below:
$$s^2(t)=N\cdot\smallint\limits_t^{\infty}h^2(x)dx$$
where $h^2$ is the squared impulse response, $N$ is the noise power per unit bandwidth and $s^2(t)$ is the energy. A linear least-squares fit is then performed on the decay curve from $-5 \dB$ to $-35 \dB$. That yields $RT_{30}$ (the time it takes for $SPL$ to drop 30 $\dB$). It is then extrapolated to $RT_{60}$ by multiplying by a factor of 2.

DRR, C50 and C80 ratios were extracted from RIR measurements using the following equations, using respectively $t= 2.5ms$, $t= 50ms$ and $t=80ms$:

$$X_{t}=10\log{\frac{\int_0^{t[ms]} h^2(\tau)d\tau}{\int_{t[ms]}^{\infty} h^2(\tau)d\tau}}[\dB]$$

where $h$ is the sound pressure level and $X_t$ the parameter.


STI was derived from the RIRs using SoundZoneTools \cite{SoundZoneTools}. The SNR of each audio sample was selected randomly from -5 to 24 $\dB$, in increments of 1 $\dB$.
In order to reach the desired SNR, the noise signal was generated at a given intensity with respect to the speech signal. 
The SNR is defined as:
$$SNR=10\log{\frac{P_{x}}{P_\text{n}}}[\dB]$$
where $P_{x}$ is the power of the speech signal and $P_\text{n}$ is the power of the background noise. 
 
All audio samples were normalized, each using min-max normalization, assuming that the min of an audio signal corresponds to silence and thus should be zero: $S_{norm} = S / max(|S|)$. Data were down-sampled to mono-channel 16kHz. The generated speech files were trimmed to 8 seconds chunks since this duration showed to be a good compromise between computation time, model complexity, and evaluation score. Mel Frequency Cepstral Coefficients (MFCC) spectrums were computed and used as inputs to the network for each signal. MFCC features were chosen since they have proven to give the best results \cite{callens2020joint}. For our experiments, we used 25ms frame size and 10ms frame steps. The MFCCs were obtained using Librosa~\cite{mcfee2015librosa}.

\subsection{Neural Networks}

Networks were trained to jointly predict the SNR and five acoustic parameters: STI, RT60, C50, C80, and DRR. Table~\ref{tab:conv-params} shows the architectures of the baseline (CNN) and our proposed model (CRNN). The baseline~\cite{looney2020joint} represents the state-of-the-art model for the joint blind estimation of acoustic parameters. The proposed CRNN outperforms the CNN baseline when jointly estimating RT60, C50, C80, DRR. Baseline state-of-the-art methods for STI and SNR estimation were also included in the evaluation.

WADA-SNR (Waveform Amplitude Distribution Analysis) \cite{SNRwada} was used as a baseline for SNR estimation. This algorithm assumes that a Gamma distribution can approximate the amplitude distribution of clean speech and that an additive noise signal is Gaussian. Based on this assumption, the SNR is estimated by examining the noise corrupted speech's amplitude distribution.

For blind STI estimation, the architecture used in~\cite{STInew} was implemented as a baseline. It consists of a deep CNN that takes as input reverberant speech signals. This network was trained using simulated RIRs combined with clean speech examples. In contrast to our work, they did not include noisy signals.

The CNN and CRNN models use the ReLU and ELU activation functions, respectively. Batch normalization, dropout, and max pooling methods are used with the Conv2d layers. To optimize the training process of both models, batch training with the size of 64, early stopping with patience set to 15, ADAM optimizer~\cite{kingma2014adam} with the learning rate $\alpha=0.001$, and the mean square error loss was implemented.


\begin{table}[htbp]
\centering
\caption{Model architectures of the baseline~\cite{looney2020joint} and  our proposed CRNN. The output classes are STI, SNR, RT60, C50, C80, and DRR. ``C`` stands for Conv2D, ``G`` for GRU, and ``D`` for Dense layers.}\vspace{9pt}
\label{tab:conv-params}
\begin{tabular}{c|c|c}
 & \multicolumn{2}{c}{(size) / (kernel size, number of filters)} \\ \hline
Model & Baseline & CRNN \\
layers & \#1.66M & \#369K \\ \hline 
1 & C(5, 256) & C(3, 64) \\ \hline
2 & C(5, 256) & C(3, 128) \\ \hline
3 & D(64) & C(3, 128) \\ \hline
4 & D(4) & C(3, 128) \\ \hline
5 & - & G(32) \\ \hline
6 & - & G(32) \\ \hline
7 & - & D(128) \\ \hline
8 & - & D(64) \\ \hline
9 & - & D(4) \\
\end{tabular}
\end{table}

\subsection{Evaluation metric}
The objective is to investigate the estimation error over the SNR and five acoustic parameters: STI. RT60, C50, C80, and DRR. The mean square error (MAE) is used to evaluate and compare the models. For $n$ estimations:

$$\text{MAE} = \sum_{i=1}^n \frac{\lvert y_{pred}-y_{true}\rvert}{n}, [s\text{ or } \dB]$$





\section{Results}
\label{sec:results}

Table \ref{tab:jaudiospeech} shows the evaluation of the joint blind estimators of the SNR and the five acoustic parameters (STI, DRR,  RT60, C50, C80). Moreover, Table \ref{tab:jaudiospeech} compares the MAE in estimating the SNR and STI using the joint estimators to their estimation using WADA-SNR \cite{SNRwada} and the STI baseline \cite{STInew}, respectively. The best estimation for each parameter is highlighted in bold. Figure \ref{fig:scatteraudio} shows the CRNN model's performance to estimate DRR, T60, C50, and C80. Figure \ref{fig:baseline} compares the performance of our proposed model and the baseline models for STI  \cite{STInew} and SNR \cite{SNRwada} estimation. All the models were tested on test speech signals. 

The proposed CRNN outperforms the rest of the models for all the parameters considered. We see that the acoustic estimates degrade with more challenging environments (e.g., smaller STI or higher T60). On the other hand, the SNR prediction degrades for lower background noise (higher SNR). The network was trained using noisy and non-noisy signals. The non-noisy signals could be confused with signals that present low noise levels, which might explain a ``worse`` prediction for high  SNR values.  

\begin{table}[htbp]
\caption{Speech MAE obtained with proposed estimators.}\vspace{9pt}
\label{tab:jaudiospeech}
\centering
\begin{tabular}{l|c|c|c|c|c|c}
    \multirow{2}{*}{Method}& SNR & STI &  DRR & T60 & C50 & C80\\
    & [$\dB$] & & [$\dB$] & [s] & [$\dB$] & [$\dB$] \\ 
    \hline
    WADA-SNR & 4.69 & - & - & - & - & - \\
    STI baseline & - & 0.091 & - & - & - & - \\
    \hline
    CNN & 2.17 & 0.036 & 2.98 & 0.24 & 6.28 & 7.20 \\
    \textbf{CRNN} & \textbf{1.98} & \textbf{0.033} & \textbf{2.91} & \textbf{0.21} & \textbf{5.95} & \textbf{6.60} \\
\end{tabular}
\end{table}

\begin{figure*}[htbp]
    \centering
 
     \includegraphics[width=1.0\linewidth]{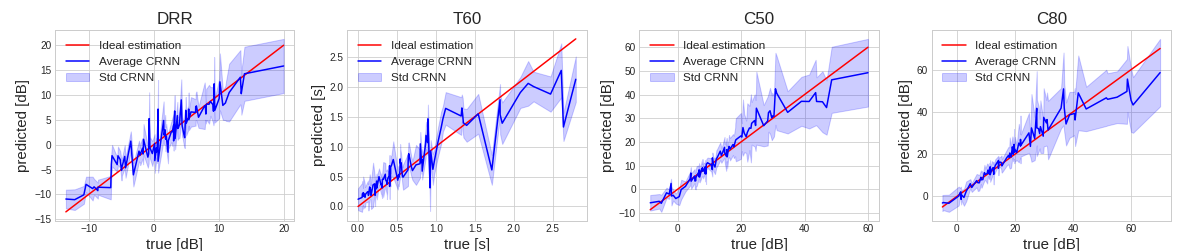}  

    \caption{CRNN model performance for joint blind estimation on a noisy and reverberant speech. The red line indicates the ideal estimation, the blue line indicates the average estimation of the model, and the blue shade indicates the standard deviation of the prediction.}
    \label{fig:scatteraudio}
\end{figure*}

\begin{figure}[htbp]
    \centering
    
    \begin{subfigure}[b]{\columnwidth}
      \centering
      \includegraphics[width=\linewidth]{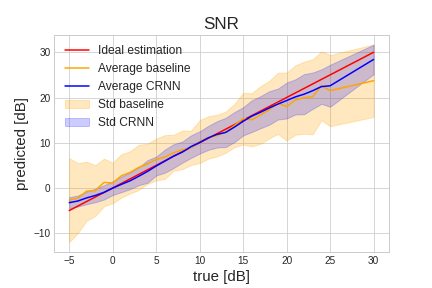}  
      \label{fig:sub-first}
    \end{subfigure}
    
    \begin{subfigure}[b]{\columnwidth}
      \centering
      \includegraphics[width=\linewidth]{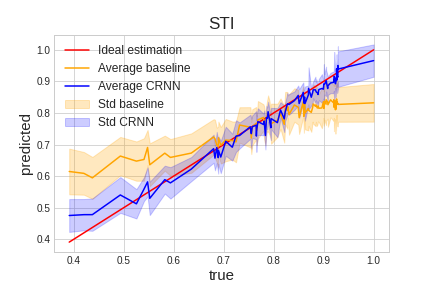}  
      \label{fig:sub-second}
    \end{subfigure}
    \caption{This figure shows a comparison of the proposed and baseline models. The top picture includes WADA-SNR baseline~\cite{SNRwada}, and the bottom one plots STI baseline~\cite{STInew}. The red lines indicate the ideal estimation. The blue lines indicate the average estimation of the  CRNN model. The yellow line indicates the average estimation of the baseline models, and the shade of the respective colors indicates the standard deviation of the prediction.}
    \label{fig:baseline}
\end{figure}

\subsection{Robustness to noise}
Table \ref{tab:noise} compare the Mean Absolute Error in estimating the acoustic parameters (STI, DRR, T60, C50 and C80) for different SNR ranges. It can be seen that  the acoustic estimates slightly degrade in environments with higher background noise. SNR values smaller than 0 $\dB$ imply that there is more noise than speech signal. This can affect the model's performance. Nevertheless, the prediction is robust for all SNR conditions.

\begin{table}[htbp]
\caption{Speech MAE on acoustic parameters estimation with CRNN for different SNR values in [$\dB$].}\vspace{9pt}
\label{tab:noise}
\centering
\begin{tabular}{l|c|c|c|c|c|}
    \multirow{2}{*}{Range SNR} & STI &  DRR & T60 & C50 & C80\\
    & & [$\dB$] & [s] & [$\dB$] & [$\dB$] \\ 
    \hline
    (-6, -1] & 0.041 & 3.68 & 0.24 & 7.93 & 8.95  \\
    (-1, 4] & 0.034 & 2.98 & 0.23 & 6.40 & 7.18 \\
    (4, 9] & 0.032 & 2.91 & 0.23 & 5.93 & 6.58 \\
    (9, 14] & 0.030 & 2.70 & 0.20 & 5.59 & 6.18 \\
    (14, 19] & 0.031 & 2.80 & 0.20 & 5.53 & 6.18 \\
    (19, 24] & 0.030 & 3.68 & 0.20 & 5.31 & 5.81 \\

\end{tabular}
\end{table}

\section{Discussion and conclusion}
\label{sec:conclusion}

In this work, we have proposed the universal acoustic environment estimator -- an end-to-end method to blindly and jointly estimate the signal-to-noise ratio (SNR) and five acoustic parameters from a noisy reverberant audio recording. This first version estimates reverberation time (RT60), direct-to-reverberant ratio (DRR), clarity (C50 and C80), and speech transmission index (STI).
The speech, noise, and RIR samples were carefully split for training and testing, and a mixture of datasets has been used. Moreover, different types of background noises were considered (real, white, and pink). This ensures the validity of the results and our algorithm's robustness against noise and unseen examples. It has been shown that the estimation of the room acoustic parameters (STI, T60, C50, and C80) is robust to different types and levels of noise. Similarly, the SNR estimation is robust to reverberant speech audio signals. We have confirmed our hypothesis that multi-task model training outperforms individual predictors for the SNR and STI parameters, and CRNN has been consistently better than CNN. 

Besides, the CRNN model has almost five times fewer parameters than the baseline CNN model. Therefore, it is more suitable for embedded platforms. The proposed algorithm could be used in an application to provide real-time feedback on speech, similarly to the proposed in~\cite{voiceAssist}. The user could benefit from this feedback to adjust the recording setup to improve the quality of voice recordings and remote presentations within others. The universal room acoustic estimator is released as a Code Ocean  \cite{codeocean} Python/TensorFlow capsule \cite{doi}.

\section{Acknowledgements}
This work was realized during a six-month internship at Logitech. 
\balance
\bibliographystyle{IEEEtran}
\bibliography{refs,references}

\end{sloppy}
\end{document}